\documentstyle[12pt,aaspp4]{article}

\def\sn{SN~1987A}

\def\hst{{\it HST\/}}
\def\Lya{${\rm Ly}\alpha$}
\def\Ha{${\rm H}\alpha$}
\def\Hb{${\rm H}\beta$}

\def\EE#1{\times 10^{\small#1}}
\def\cm3{\rm ~cm^{\small -3}}
\def\kms{\rm ~km~s^{\small -1}}

\def\ergcms{\rm\,erg~cm^{\small -2}~s^{\small -1}}
\def\ergcmsA{\rm\,erg~cm^{\small -2}~s^{\small -1}~\AA^{\small -1}}
\def\ergcmsarcs{\rm\,erg~cm^{\small -2}~s^{\small -1}~arcsec^{\small -2}}
\def\EBV{{\rm E}(B-V)}
\def\wl{$\lambda$}
\def\wll{$\lambda\lambda$}

\def\HeII{{\ion{He}{2}\,}}

\def\NII{{\ion{N}{2}}}

\def\NIV{{\ion{N}{4}}}
\def\NIVb{{\ion{N}{4}]\,}}
\def\NV{{\ion{N}{5}\,}}

\def\OI{{\ion{O}{1}}}
\def\OII{{\ion{O}{2}}}

\def\OIII{{\ion{O}{3}}}

\def\OIV{{\ion{O}{4}}}
\def\OIVb{{\ion{O}{4}]\,}}

\def\SII{{\ion{S}{2}}}

\lefthead{Sonneborn et al.}
\righthead{STIS Spectroscopy of SN 1987A}

\begin{document}

\title{Spatially Resolved STIS Spectroscopy of SN 1987A: Evidence for Shock
Interaction with Circumstellar Gas\altaffilmark{1}}

\author{G. Sonneborn,\altaffilmark{2}
C. S. J. Pun,\altaffilmark{2,3}
R. A. Kimble,\altaffilmark{2,4}
T. R. Gull,\altaffilmark{2,4}
P. Lundqvist,\altaffilmark{5}
R. McCray,\altaffilmark{6}\\
P. Plait,\altaffilmark{7}
A. Boggess,\altaffilmark{4,8}
C. W. Bowers,\altaffilmark{2,4}
A. C. Danks,\altaffilmark{4,9}
J. Grady,\altaffilmark{10}
S. R. Heap,\altaffilmark{2,4}\\
S. Kraemer,\altaffilmark{4,11}
D. Lindler,\altaffilmark{7}
J. Loiacono,\altaffilmark{10}
S. P. Maran,\altaffilmark{4,12}
H. W. Moos,\altaffilmark{4,13} and\\
B. E. Woodgate\altaffilmark{2,4}}
\affil{}

\altaffiltext{1}{Based on observations with the NASA/ESA {\it Hubble
Space Telescope}, obtained at the Space Telescope Science Institute,
which is operated by the Association of Universities for Research in
Astronomy Inc., under NASA Contract NAS5-26555.}
\altaffiltext{2}{Laboratory for Astronomy and Solar Physics, Code 681,
NASA Goddard Space Flight Center, Greenbelt, MD 20771; ~sonneborn,
kimble, gull, bowers, hrsheap, woodgate @stars.gsfc.nasa.gov}
\altaffiltext{3}{National Optical Astronomy Observatories, Tucson, AZ
85726; ~pun@congee.gsfc.nasa.gov}
\altaffiltext{4}{Space Telescope Imaging Spectrograph Investigation Definition
Team}
\altaffiltext{5}{Stockholm Observatory, S-133 36, Saltsj\"obaden, Sweden;
~peter@astro.su.se}
\altaffiltext{6}{JILA, University of Colorado, Boulder, CO 80309;
~dick@jila.colorado.edu}
\altaffiltext{7}{Applied Computer Concepts, Code 681, NASA Goddard
Space Flight Center, Greenbelt, MD 20771; plait@abba.gsfc.nasa.gov,
lindler@rockit.gsfc.nasa.gov}
\altaffiltext{8}{2420 Balsam Drive, Boulder, CO 80304; 
boggess@lyrae.colorado.edu}
\altaffiltext{9}{Hughes STX, Code 681, NASA Goddard Space Flight
Center, Greenbelt, MD 20771; ~danks@stars.gsfc.nasa.gov}
\altaffiltext{10}{Laboratory for Astronomy and Solar Physics, Code 680.1,
NASA Goddard Space Flight Center, Greenbelt, MD 20771; ~grady,
loiacono@stars.gsfc.nasa.gov}
\altaffiltext{11}{Catholic University of America,
Code 681, NASA Goddard Space Flight Center, Greenbelt, MD 20771;  
kraemer@stars.gsfc.nasa.gov}
\altaffiltext{12}{Space Sciences Directorate, Code 600, NASA Goddard Space
Flight Center, Greenbelt, MD 20771; ~hrsmaran@eclair.gsfc.nasa.gov}
\altaffiltext{13}{Dept.\ of Physics and Astronomy, Johns Hopkins
University, Baltimore, MD 21218; ~hwm@pha.jhu.edu}

\begin{abstract}

Visual and ultraviolet spatially resolved $(\sim0\farcs1)$ spectra of 
\sn\ obtained on days 3715 and 3743 with the Space Telescope Imaging 
Spectrograph on the Hubble Space Telescope show that the high-velocity 
SN debris is colliding with circumstellar gas.  Very broad \Lya\ 
emission with velocities extending to $\sim\pm20,000 \kms$ originates 
{\it inside} the inner circumstellar ring and appears to fill most of 
the surface area within 0\farcs67 $\pm$ 0\farcs03 (0.14 pc at a 
distance of 50 kpc) of the ring's center.  The observed \Lya\ flux 
from the shocked ejecta is $(1.85 \pm 0.53)\EE{-13} \ergcms$ and 
$(1.25 \pm 0.51)\EE{-12} \ergcms$ after correcting for extinction.  A 
spatially {\it unresolved} blue-shifted emission feature was 
discovered in \Ha\ (and other lines) on the inner ring at 
p.a.\,$31\arcdeg \pm 8\arcdeg$.  The \Ha\ emission extends to $-250 
\kms$ with no corresponding red-shifted emission.  This highly 
localized interaction appears to be the initial contact of the 
supernova blast wave with an inward protrusion of the inner ring.  The 
broad \Lya\ emission and the `hot spot' are separate interaction 
phemonena associated with the reverse and forward shocks, 
respectively.  We also find that the size of the inner ring in 
forbidden lines of oxygen has a dependence on ionization potential, in 
agreement with photoionization models of the ring.

\end{abstract}

\keywords{supernovae: individual(\sn) -- supernova remnants --
circumstellar matter}

\section{INTRODUCTION}

\sn\ in the Large Magellanic Cloud continues to provide new
opportunities to study the late-time evolution of supernovae 
and to probe their stellar origins.  The collision of the 
ejecta with the progenitor's circumstellar (CS) medium has been 
anticipated since its discovery in mid-1987, due to its proximity to 
the supernova (Fransson et al.\ 1989).  The coincident emergence of 
radio (Staveley-Smith et al.\ 1992, 1993) and X-ray (Beuermann et al.\ 
1994) emission in 1990, their subsequent monotonic increase, and the 
deceleration of the supernova shock (Staveley-Smith et al.\ 1993) 
indicate that the high-velocity SN shock encountered low-density CS 
gas starting around 1200 days after the explosion.  This gas 
corresponds to mass lost by the progenitor near the end of its red 
supergiant (RSG) phase (Chevalier \& Dwarkadas 1995, hereafter CD95).  
Gaensler et al.\ (1997) find the radio emitting region is located 
just inside the central ring, based on a $\sim$0\farcs4 resolution 
9~GHz map, and is expanding at $\sim 2,800 \kms$ (12 mas yr$^{-1}$).  The 
shock interaction region, bounded by 
forward and reverse shocks, consists of an outer zone of shocked CS gas 
and an inner zone of shocked ejecta (Chevalier 1982).

The \sn\ CS medium is a fossil record of the pre-explosion mass loss 
and may hold important clues to the nature of the progenitor (e.g.\  
single vs.  binary star).  The gas in the central ring ($n_e\sim 
10^{4}$ cm$^{-3}$) is significantly enhanced in helium and nitrogen, 
indicating that the progenitor was in a post He-core burning phase when 
it exploded (Fransson et al.\ 1989, Sonneborn et al.\ 1997).  The 
presence of lower density gas ($n_e\sim 100$ cm$^{-3}$, CD95) inside 
the ring traces the mass loss at the end of the RSG stage, after 
formation of the inner ring.

The small angular extent and complex structure of the \sn\ CS gas and 
debris require the high spatial resolution of the Hubble Space 
Telescope (\hst).  WFPC2 and FOC imagery revealed the remarkable 
thin-ring structures (Jakobsen et al.\ 1991, [\OIII]; Burrows et al.\  
1995, \Ha; Plait et al.\ 1995, [\NII]).  However, the first generation 
\hst\ spectrographs had one-dimensional detectors and could only 
provide spatial information on the scale of the available apertures.  
The Space Telescope Imaging Spectrograph (STIS) provides imaging 
spectroscopy by combining excellent optical performance of imaging 
optics and two-dimensional detectors (Kimble et al.\ 1997).  With good 
spectral resolution and \hst-limited spatial resolution, we are now 
able to study the spatial dependence of the line emission from the 
inner ring and do the first spectral analysis of the interaction of 
the SN ejecta with the CS gas.  In an accompanying paper, Michael et 
al.\ (1997, hereafter M97) model the CS interaction and the emission 
processes that produce the \Lya\ emission discussed below.

\section{OBSERVATIONS}

\subsection{Ultraviolet Spectral Imaging}

Ultraviolet spectral images of \sn\ and the inner CS ring were taken 
with the STIS FUV-MAMA detector on 1997 May 24 (3743 days after core 
collapse).  Two exposures (Pgm ID 7123) were taken in 
time-tag mode with the G140L grating ($\sim 1130 - 1720$~\AA) for a 
total exposure time of 4180 sec.  The plate scale of the image is 
0\farcs0244 per pixel in both the spatial and spectral directions.  
The spatial resolution is $\sim0\farcs06$.  The G140L spectral resolution
is $\sim$~1.2~\AA\ (0.5833 \AA/pixel).  The $2\arcsec 
\times 2\arcsec$ aperture was chosen to include the entire inner 
circumstellar ring ($1\farcs7 \times 1\farcs2$, Plait et al.\ 1995) 
and exclude the neighboring B stars, primarily star~3 ($V \sim 
16$, 1\farcs6 from the SN; see Fig.~1a).  The G140L spectral image is 
shown in Figure~2.

A striking and previously unobserved emission feature, extending from 
$\lesssim1145$ to $\sim1300$\,\AA, is identified as \Lya\ from the 
shock interaction of the high-velocity debris with CS gas.  The strong 
radiation filling the square aperture is geocoronal \Lya.  The diffuse 
continuum extending the length of the spectrum is UV background from 
the LMC.  The detector background count rate is extremely low 
($<1\EE{-5}$ cts~s$^{-1}$~px$^{-1}$, Kimble et al.\ 1997) 
and is negligible in this data.  Images of the inner ring are also 
detected in \OIV]\,\wll1397--1407, \NIV]\,\wl1486.5, and 
\HeII\,\wl1640.5;  their orientation  
indicates that the aperture's spatial axis is at p.a.\,$115\arcdeg \pm 
4\arcdeg$.  A weak, spatially extended UV continuum (1400 -- 1700~\AA) 
from the debris is also present.

Wavelength calibration exposures using the Pt-Cr/Ne line lamp were 
taken through the $52\arcsec \times 0\farcs1$ slit for each G140L 
exposure.  The absolute wavelength scale was referenced to the small 
offset ($\Delta x$) of the SN from the center of the aperture.  We 
find the SN offset to be $\Delta x$ = +0\farcs12 $\pm$ 0\farcs15 in 
the dispersion direction and $\Delta y$ = $-$0\farcs17 $\pm$ 0\farcs04 
in the spatial direction, as determined from the center of the ring in 
\OIVb, \NIVb, and \HeII.

\subsection{Visible Spectral Imaging}

Medium-resolution spectral images (Pgm ID 7123) of the inner ring were 
obtained in two grating settings on 1997 April 26 (day 3715): G750M 
(6295~-- 6867~\AA, resolution = 1.02~\AA, 643~sec), and G430M (4818~-- 
5104~\AA, resolution = 0.83~\AA, 828~sec).  The $52\arcsec \times 
2\arcsec$ slit was chosen to include the entire ring and was oriented along  
its major axis 
(Fig.~1b), thereby providing maximum separation in the dispersion direction 
between images of the ring in adjacent emission lines.  The narrow 
emission lines from the ring ($v_{\small \rm FWHM}\lesssim 15 \kms$) 
remain spectrally unresolved in these observing modes ($\Delta v \sim 
50 \kms$), while the debris spectral features ($v_{\small \rm 
FWHM}\sim 2500 \kms$) are highly dispersed.  In addition to the CS 
ring and the SN debris, the G750M and G430M spectral images (see 
Figure~3) also show diffuse \Ha, \Hb, and [\OIII] emission filling the 
long slit and the continuous spectra of star 3 (including stellar \Ha\ 
and \Hb\ emission lines) and several fainter stars.

An [\OII] \wl3727 image (Pgm ID 7122) of \sn\ and its ring system was 
taken with STIS on 1997 April 4 (day 3693), using the F28X50OII 
filter.  Four exposures totaling 2415~sec were obtained.

\section{RESULTS}

\subsection{Ultraviolet Spectroscopy}

The \Lya\ emission from \sn\ reported here is the signature of the 
shock interaction between high-velocity SN ejecta and CS gas inside 
the bright inner ring.  Two aspects of the \Lya\ feature are 
particularly noteworthy: a) the spatial and velocity distribution of 
the \Lya\ flux, and b) the relative brightnesses of the red and blue 
wings of the line profile.

We measure the red edge of the \Lya\ profile at $1301.7 \pm 3.7$~\AA, 
or $+20,900 \pm 900 \kms$ relative to 1216.825~\AA, \Lya\ in the LMC 
rest frame.  The red edge could, however, be as large as 1328~\AA\ 
($27,400 \kms$), depending on the details of the diffuse background 
subtraction.  A Doppler shift of $-20,900 \kms$ on the blue side 
of the profile occurs 
at 1131~\AA.  Unfortunately, we cannot measure the profile shortward 
of $\sim1145$~\AA\ due to decreasing \hst/STIS sensitivity below 
1170~\AA.  Measurement of the blue wing of the \Lya\ profile awaits 
observations by the Far Ultraviolet Spectroscopic Explorer (FUSE) in 
1999.

The \Lya\ brightness profile in the spatial direction (across the 
spectrum) sets a limit on the spatial extent of the shock interaction 
region in a direction close to the major axis of the ring.  Figure 4 
shows spatial brightness profiles in the blue and red wings of \Lya.  
The \Lya\ surface brightness is fairly uniform and there is no 
significant brightening along the upper and lower edges of the feature 
(cf.\ M97).  The spatial extent is 1\farcs22 $\pm 0\farcs06$ and 
1\farcs14 $\pm$ 0\farcs11 on the red ($1250\pm8$ \AA) and blue 
($1179\pm8$ \AA) sides of the aperture.  The projected size of the 
ring in the observing geometry is 1\farcs51 $\pm$ 0\farcs06, based on 
the ring size in [\OIII]  (Plait et al.\ 1995).  Thus the 
interaction region is located at 79\% $\pm$ 4\% of the ring's radius, 
or $\sim5\EE{17}$ cm from the center of the debris.  The spatial 
extent of the interaction region in the dispersion direction cannot be 
measured directly from the STIS data, due to the fact that the 
observed $(x,y)$ distribution of \Lya\ photons is a convolution of the 
geometry and the velocity field.  The separation of the two effects is 
model-dependent (M97), but the spatial extent is probably 
$\lesssim1\farcs4$.  The blue and red wings are tapered toward higher 
velocities and the maximum Doppler shift is roughly aligned with the 
center of the ring.

The \Lya\ line profile (Fig.\ 5) is the sum of the observed count rate 
at each wavelength point multiplied by the G140L sensitivity function.  
The integrated observed flux assumes that the part of the \Lya\ 
profile obscured by geocoronal emission decreases linearly from 1187.0 
\AA\ to 1241.3 \AA\ (see Fig.~5), corresponding to the edges of the 
2\arcsec-wide aperture in the G140L image.  The points where the 
profile meets the background are 1144.4 and 1301.7 \AA.  The 
integrated \Lya\ flux is $(1.85 \pm 0.53) \EE{-13} \ergcms$, 
consistent with the hydrodynamical models of Borkowski, Blondin, \& 
McCray (1997).  The dereddened \Lya\ flux is $(1.25 \pm 0.51) \EE{-12} 
\ergcms$ ($\EBV=0.16$, Fitzpatrick \& Walborn 1990).  The dereddened 
fluxes from the observable portions of the profile are $(3.3 \pm 0.9) 
\EE{-13} \ergcms$ (1144.4~-- 1187.0~\AA) and $(2.3 \pm 0.5) \EE{-13} 
\ergcms$ (1241.3~-- 1301.7~\AA).  This asymmetry is probably due to 
the $\sim$10-month light travel time across the interaction region, 
implying a short time scale for changes in the shock spectrum.  Dust 
in the ejecta might also be a factor.  There is no conclusive evidence 
for \NV\,\wl1240 emission from the shocked gas (see M97).  The 
intensity variations in the red wing may be the result of strong 
interstellar absorption lines (e.g., 1260, 1300, and 1335~\AA, cf.\ 
Sonneborn et al.\ 1997) convolved with the spatial extent of the 
interaction region.

There are significant uncertainties in the \Lya\ flux measurements.  
The net effect of statistical errors is very small ($<1$\%) and is 
negligible by comparison with the other error sources: the geometry of 
the interaction region, systematic errors in the flux measurement, 
interstellar extinction, and the \hst/STIS G140L sensitivity function.

We estimate that the uncertainty in the evaluation of the \Lya\ flux 
hidden by geocoronal emission (see Fig.\ 5) is $\sim$25\%.  The 
estimated uncertainty in the preliminary G140L sensitivity function 
used for this flux calculation is $\sim 20$\% longward of $\sim 
1240$\,\AA\ and increases to $\sim 40$\% at the shortest wavelengths.  
(The post-launch STIS calibration is currently underway and the 
expected uncertainty for the complete sensitivity function should be 
$\lesssim 10\%$ for all wavelengths).  Other potential sources of 
systematic error include the selection of the end points of the 
integration regions and background subtraction.

The  spatial extent of the CS interaction region in the 
dispersion direction $(\pm0\farcs7= \pm 16.7\, {\rm\AA} = \pm 4100 
\kms)$ introduces uncertainties in the calibrated \Lya\ profile, 
unless its geometry is known and modelled.  If the \Lya\ surface 
brightness is symmetric, as may be the case for \sn\ (M97), this 
effect would be significantly reduced.  Nevertheless, we estimate this 
uncertainty to be 20~-- 25\% of the \Lya\ flux.

Errors in the Galactic and LMC extinction laws, in particular at 
$\lambda<1200$ \AA, are an additional source of error in the 
dereddened flux.  The uncertainty in the extinction law near 1200~\AA\ 
is about 10\% (Fitzpatrick \& Walborn 1990), while for 1140~-- 
1200~\AA\ we estimate it to be $\sim 20$\%.  The uncertainty in E(\bv) 
for \sn\ is 0.02~mag (Fitzpatrick \&~Walborn 1990).  The net 
uncertainty due to extinction is $\sim$~25\%.  Combining these error 
sources, the total estimated 
uncertainty in the observed \Lya\ flux is 30~-- 35\% (38~-- 43\% for 
the dereddened flux).

Apart from the \Lya\ emission, the inner ring is detected in the G140L 
image in \OIV]\,\wll1397--1407, \NIV]\,\wl1486.5, and \HeII\,\wl1640.5 
(see Fig.~2), indicating that a significant fraction of hot gas 
remains in the ring.  The dereddened surface brightnesses for \OIV], 
\NIV], and \HeII\ are $(6.0 \pm 1.4)$, $(10.8 \pm 2.3)$, and $(15.4 
\pm 3.4)\EE{-15} \ \ergcmsarcs$, respectively.  In addition, a faint 
and spatially extended far-UV continuum of the SN debris $(FWZI 
\sim0\farcs44$ and centered on the \OIV], \NIV], and \HeII\ images of 
the ring) is detected longward of 1400 \AA\ with an average dereddened 
flux of $(1.6 \pm 0.6)\EE{-16} \ergcmsA$.  Although a $V\sim 20$ A5 
star (Plait et al.\ 1995) is superposed on the ring at 
p.a.\,227\arcdeg\ (see Fig.\ 1),  such a star would contribute less 
than 5\% of the observed flux. There is no evidence for a point 
source spectrum in the G140L data.

\subsection{Visible Spectroscopy}

The STIS G430M and G750M observations of \sn\ yield nearly 
monochromatic images of the inner CS ring in a large number of lines: 
\Hb, [\OIII]\ \wll4958.9, 5006.8, [\OI]\ \wll6300.3, 6363.8, [\NII]\ 
\wll6548.0, 6583.4, \Ha, and [\SII]\ \wll 6716.5, 6730.8, are 
observed.  Parts of the outer rings are noticeable in \Ha, [\NII], and 
[\OIII].  \Ha\ emission from the SN debris is detected as a very 
broad, spatially resolved emission feature passing through the center 
of the [\NII] and \Ha\ ring images.  The observed images of the ring 
(Fig.~6) have similar morphology and intensity variations.

These spectral images revealed a spatially {\it unresolved}, and 
previously unknown, 
blue-shifted \Ha\ emission feature (also detected in \Hb, 
[\OI]\,\wl6300, and [\OIII]\,\wl5007) located close to or on the inner 
ring at p.a.\ $31\arcdeg \pm 8\arcdeg$ (Pun et al.\ 1997).  The \Ha\ 
emission extends to $-250 \kms$ with no corresponding red-shifted 
emission (see Fig.  6d).  Garnavich, Kirshner, \& Challis (1997) 
pinpointed this `hot spot' when they found a brightening at the same 
p.a., but located slightly inside the ring, in 1997 July WFPC2 \Ha, 
[\OIII], and broad-band images.  The hot spot must be a separate 
phenomenon from that producing the spatially-extended, high-velocity 
\Lya\ emission.  It may be a manifestation of the initial encounter of 
the SN blast wave with an inward protrusion of the inner ring.  If so, 
we would expect it to continue to brighten rapidly and that more such 
spots will appear during the next few years.

The STIS observations also help determine the location of the outer 
rings relative to the SN.  A portion of the northern outer ring  
is projected on the SN debris (cf.\ Fig.\ 1).  In Fig.\ 6d, the image of 
this ring in [\NII]\,\wl6583 has a $\sim0\farcs3$ gap where it crosses 
the SN position, indicating 
that this section is located behind the debris.

The size of the ring as a function of p.a.\ in [\OI]\,\wl6300, 
[\OII]\,\wl3727, [\OIII]\,\wl5007, and [\SII]\,\wl6731 is shown in 
Figure 7.  In most places the ring size in the oxygen lines increases 
with ionization stage.  The radius is most reliably measured at the E 
ansa (p.a.\,88\arcdeg), where $\theta = 0\farcs792$ and 0\farcs845 for 
[\OI] and [\OIII], respectively.  The [\SII]\,\wl6716/\wl6731 and 
[\NII]\,\wl6548/\wl6583 line ratios (Fig.~8) are more uniform than 
the individual line images (Fig.\ 6).  The mean [\SII] ratio is $0.54 
\pm 0.02$, corresponding to $N_e =$ 6000 cm$^{-3}$.  This is similar 
to that found by Lundqvist et al.\ (1997), based on an analysis of \hst/FOS 
ring spectra from earlier epochs at p.a.\  $\sim$300\arcdeg.  The 
presence of a significant ionization gradient in the ring and a 
fairly uniform density for the emitting gas are consistent with the 
photoionization/recombination models (Lundqvist \& Fransson 1996; 
Lundqvist \& Sonneborn 1997).

\section{CONCLUSIONS}

STIS UV spectral imaging has further characterized the shock 
interaction region around \sn.  Broad \Lya\ emission originates {\it 
inside} the inner ring, at $\sim80$\% of the ring's radius, 
corresponding to the location of the reverse shock front (M97).  The 
fairly uniform \Lya\ surface brightness and large Doppler shifts 
indicate that the shock geometry is not confined to the plane of the 
inner ring and may in fact be spheroidal, as shown by M97.  This may 
provide evidence about the geometry of the RSG wind above and below 
the ring plane.  The emerging picture of the debris-CS gas interaction 
region is close indeed to that proposed by CD95.  The asymmetry 
between the blue and red wings of \Lya\ indicates that the emission 
from the reverse shock may be evolving on a time scale of months.  The 
location of the reverse shock is consistent with unimpeded debris 
expansion at $\sim 40,000 \kms$ up to $\sim1200$ days, when the radio 
emission picked up, followed by expansion of the shock at $\sim 2,800 
\kms$ to the present.

The hot spot discovered at p.a.\,31\arcdeg\ appears to be a different, 
highly localized shock interaction phenomenon closer to the ring than 
that observed in \Lya.  The location of the hot spot indicates that 
the forward shock has reached the innermost edge, or an inward 
protrusion, of dense ring material.  This places the forward shock at 
$\sim 1.1$ times the the reverse shock radius.  If this picture is 
correct, similar hot spots may light up around the ring in the coming 
years, providing a detailed map of the ring's radial geometry.

\acknowledgments

We are grateful to Bob Kirshner and the SINS collaboration for sharing 
their recent WFPC2 results with us before publication and to Roger 
Chevalier and Claes Fransson for useful discussions.  We thank the 
staff of the Space Telescope Science Institute and the STIS Team at 
GSFC and Ball Aerospace for their support and the
\hst\ Project for the opportunity to make these observations as part 
of the Early Release Observation program.

\clearpage

\begin{center}
Figure Legends
\end{center}

FIGURE 1. --- The size and location of the STIS apertures are shown 
on a 1995 composite WFPC2 image of \sn\ (courtesy of P. Challis, R. 
Kirshner, and the SINS collaboration). The image size is 13\arcsec\ 
square. a) The 2\arcsec $\times$ 
2\arcsec\ aperture location for 
the G140L observation. b) The 52\arcsec $\times$ 2\arcsec\ aperture
orientation (p.a.\,87\arcdeg) for the G430M and G750M exposures.

FIGURE 2.  --- a) STIS FUV-MAMA G140L (1130 -- 1720 \AA) raw spectral 
image of \sn.  The image was reconstructed from time-tag mode data.  
The displayed image is a $1024 \times 300$ pixel subarray of the 1024 $\times$ 
1024 image.  Although star 3 lies outside the aperture, some light 
from it is still detected in the G140L image as the continuum emission 
along the top edge of the spectrum.  b) Expanded view of Fig.~2a, with 
background corrections.  
Flux from star~3 was subtracted with row-by-row continuum 
fits to the data.  The diffuse background  was approximated by 
extracting the spectrum in the $\sim$~0\farcs2 gap between the 
extended \Lya \ emission and the light from star~3.  This spectrum was 
subtracted from every row of the image after it was scaled to match 
the gradient in the background along the slit.  Both images are 
displayed on a logarithmic intensity scale.  Wavelength 
scales are shown in the LMC rest frame.

FIGURE 3.  --- a) STIS G430M 4851\AA\ spectral image of \sn\ shows the 
images of the inner circumstellar ring in \Hb\ and [\OIII]\,\wl4959 
and \wl5007.  These lines are also present in diffuse emission from 
the LMC filling the $52\arcsec \times 2\arcsec$ aperture.  The bright 
point source spectrum is from star~3.  The image shown is a $950 
\times 450$ pixel subarray of the 1024 $\times$ 1024 image.  The image 
was processed to remove cosmic rays and hot pixels.  b) same as 
Fig.~3a, for the G750M 6581\AA\ exposure.  The images are shown on a 
logarithmic intensity scale and the contrast is stretched to enhance 
weaker features.  The spatial scale in the dispersion and 
cross-dispersion directions are not identical in the raw G430M and 
G750M images due to anamorphic distortion, a normal characteristic of 
gratings used in the off-Littrow mode.  The corrected images 
shown here are stretched in the dispersion direction by factors of 
1.0776 (G750M/6581~\AA) and 1.1223 (G430M/4961~\AA) to achieve 
0\farcs0507 $\times$ 0\farcs0507 pixels.  Wavelength 
scales are shown in the LMC rest frame.

FIGURE 4.  --- Spatial distribution of the \Lya\ emission along the 
aperture.  The spatial scale goes from the lower edge (0\arcsec) to 
the top edge (2\arcsec).  The cross-cut in the red wing was made at 
$1250\pm8$~\AA\ and at $1179\pm8$~\AA\ in the blue wing.  The vertical 
dashed lines mark the location of the inner edge of the CS ring.

FIGURE 5.  --- Calibrated STIS G140L spectrum of \sn.  The thick line 
indicates where the \Lya\ profile was measured for the integrated 
flux.  The CS \Lya\ emission within the aperture is approximated by 
the broken line. The feature at 1470 \AA\ is a combination of \NIV] 
and SN debris.

FIGURE 6.  --- STIS images of the \sn\ inner CS ring.  a) [\SII]\,\wl6716 
and \wl6731.  b) [\OII]\,\wl3727.  c) [\OIII]\,\wl5007.  d) 
[\NII]\,\wl6548, \Ha, [\NII]\,\wl6583.  e) \Hb.  f) [\OI]\,\wl6300.  
Images a), c), d), e), and f) are subarrays of the images shown in 
Fig.\ 3.  Wavelength increases from left to right.  The [\OII] direct 
image (b) was taken with the F28X50OII filter; the \sn\ debris and 
star 3 are visible (cf.\  Fig.\ 1).  
The  pixel size of these images is 0\farcs05 $\times$ 0\farcs05.

FIGURE 7.  --- Size of the \sn\ inner CS ring as a function 
of position angle in the light of several emission lines.

FIGURE 8.  --- Brightness ratios [\SII]\,\wl6716/\wl6731 (right) and 
[\NII]\,\wl6548/\wl6583 (left) for the \sn\ inner ring from the G750M 
exposure.  For the [\NII] ratio, only points with errors $<0.25$  are 
included, a total of 373 points.

\end{document}